# Clinical Trial Information Extraction with BERT

Xiong Liu[1], Greg L. Hersch[2], Iya Khalil[1], Murthy Devarakonda[1]
[1]Data Science and AI, Novartis, Cambridge, MA, USA
[2]Global Drug Development, Novartis Pharma AG, Basel, Switzerland
xiong.liu at novartis dot com

*Abstract*—**Natural language processing (NLP) of clinical trial documents can be useful in new trial design. Here we identify entity types relevant to clinical trial design and propose a framework called CT-BERT for information extraction from clinical trial text. We trained named entity recognition (NER) models to extract eligibility criteria entities by fine-tuning a set of pre-trained BERT models. We then compared the performance of CT-BERT with recent baseline methods including attention-based BiLSTM and Criteria2Query. The results demonstrate the superiority of CT-BERT in clinical trial NLP.**

## I. Introduction

Clinical trial designs are documented in unstructured text and natural language processing (NLP) of the documents can inform new trial design [1], [2]. The key NLP task is information extraction, including named entity recognition (NER) and relation extraction from various sections of the clinical trial document, such as objectives, outcomes, and eligibility criteria.

Clinical trial NER requires a fine-grained entity type system and large-scale annotation data in order to generate high quality models that meet the specific requirements of clinical trial design. Traditional clinical NER based on clinical text and the standard problem-treatment-test annotation system may not be sufficient to capture critical distinctions such as allergy, consent, language fluency and technology access.

A number of NER methods have been developed for clinical trial parsing, including rule-based methods [3] and machine learning-based methods [4]. More recently, deep learning-based methods, such as attention-based BiLSTM (Att-BiLSTM) [5], [6], have been introduced to enable more automatic and accurate extraction of entities from the eligibility criteria. Meanwhile, pre-trained language models, such as BERT, have demonstrated superior performance over previous baselines across NLP tasks [7], [8]. Investigating transformers for better clinical trial NER is a promising area of research.

We propose a new framework, called CT-BERT, to train and evaluate information extraction models based upon publicly available clinical trials data and BERT embeddings. In this work, we focus on the NER task to extract entities from the eligibility criteria. We leverage pre-annotated ClinicalTrials.gov data [5] to fine-tune a set of pre-trained BERT models for clinical trial NER, including standard BERT [7], BioBERT [8], ClinicalBERT [9] and BlueBERT [10]. To evaluate extraction quality, we use the benchmark dataset described in [4] to measure the NER performance of CT-BERT, as well as baseline models including Att-BiLSTM [5] and the conditional random field (CRF) model used in Criteria2Query [4].

Our contributions include: 1) we introduce a comparative framework for clinical trial NER, which enables the building of BERT-based models and the comparative study with other baseline models; 2) we provide empirical results for the impact of BERT pre-training on the performance of clinical trial NER; 3) we show that BERT-based NER models fine-turned on ClinicalTrials.gov data outperform baseline models.

## II. Materials and Methods

### A. Clinical trial entity type system

The rationale for clinical trial NER is that the entity types must allow us to capture the key variables (entities) in trial design. For eligibility criteria, we identified that the annotation schema in [5] is comprehensive and therefore selected it for type definition. It includes 15 types, covering common types in clinical text (e.g., disease, treatment, clinical variable) as well as speciality types (e.g., consent, language frequency, technology access) and value ranges (e.g., lower and upper bound). We treat the lower and upper bound entities as "attribute" entities because they are modifiers of other entities. To enable comparative study with Criteria2Query, we map the attribute entities and other 13 regular entities to the corresponding ones in Criteria2Query. Table I shows the detailed mapping of entities and attributes.

### B. CT-BERT architecture

The CT-BERT architecture includes both NER and relation extraction. The NER module is based on the BERT architecture. The input is clinical trial text. BERT pre-trained transformer is used as the embedding and encoding layer. And the standard BIO tag prediction in BERT is used for entity extraction. The relation extraction module is used to associate attribute entities with their base entities, which will be discussed in the extended version of the paper.

We wanted to experiment with a wide range of pre-trained models, covering a wide-range from the general domain to scientific articles (PubMed) and clinical documents (MIMIC III) so that we can assess the impact of pre-training corpora on information extraction from clinical trials documents. Therefore, we used publicly available pre-trained models, including BERT (trained on general corpus), BioBERT (trained on PubMed/PMC), ClinicalBERT (trained on MIMIC), and BlueBERT (trained on PubMed or PubMed + MIMIC).

TABLE I. ENTITIES AND ATTRIBUTES IN CT-BERT AND MAPPING TO CRITERIA2QUERY

|  | CT-BERT Class | Criteria2Query Class | Example |
|---|---|---|---|
| Entity | Age | Demographic | Age |
|  | Allergy name | Condition | Hypersensitivity to nivolumab |
|  | BMI | Measurement | Body mass index |
|  | Cancer | Condition | Thymoma |
|  | Chronic disease | Condition | Inflammatory disease |
|  | Clinical variable | Measurement | Arterial Oxygen Saturation |
|  | Contraception consent | Not found | not willing to use double barrier methods of contraception |
|  | Ethnicity | Observation | Residents of Puerto Rico |
|  | Gender | Demographic | Men and women |
|  | Language fluency | Not found | Unable to read |
|  | Pregnancy | Condition | Pregnancy |
|  | Technology access | Not found | Electronic patient diary |
|  | Treatment | Procedure | drug | Solid organ transplantation | immune-checkpoint inhibitors |
| Attribute | Lower bound | Value | Age over 18 years |
|  | Upper bound | Value | daily opioid dose <= 160 mg/day |

We adapted the dataset in [5] for NER fine-tuning. The original data uses its own off-set representation schema. We transformed the dataset into the standard BIO representation schema. The training data includes 102,985 entities in 40,876 criteria sentences and the test data includes 11,317 entities in 4,482 criteria sentences. We used the training data to fine tune the pre-trained BERT models and used the test data to measure the model performance.

## C. Evaluation of methods

To test CT-BERT NER models, we used a publicly available benchmark from [4]. It includes 10 clinical trials and 125 criteria sentences randomly sampled from ClinicalTrials.gov. The same 10-trial evaluation data has been used to evaluate Att-BiLSTM and Criteria2Query. We used the precision, recall and F1 metrics described in the Att-BiLSTM [5] and Criteria2Query [4] work. The studies employed exact match for entity type but a partial match for the entity span.

## III. RESULTS

As in previous studies, we also observed that the pre-training corpora have an impact on the NER performance, with certain additional nuances. BioBERT large pre-trained on PubMed and PMC achieves the best F1 score of 0.781 measured on the test data. The models pre-trained on MIMIC achieve an average F1 of 0.773. While the models pre-trained on the PubMed + MIMIC corpora have the lowest average F1 of 0.768. This confirms our intuition that clinical trials look more like published papers than clinical notes. And the combination of PubMed and MIMIC may have introduced biased contexts that are not representative of clinical trial documents.

We compared our fine-tuned BioBERT large with Att-BiLSTM and Criteria2Query using the benchmark data. Table II shows that the F1 scores are 0.844, 0.802, and 0.804, for the models respectively. This shows the benefit of using BERT-based models in clinical trial information extraction.

TABLE II. PERFORMANCE OF NER MODELS ON THE 10-TRIAL EVALUATION DATASET

| Evaluation | Precision | Recall | F1 |
|---|---|---|---|
| CT-BERT model (fine-tuned on BioBERT large) | 0.953 (163/171) | 0.758 (163/215) | 0.844 |
| Att-BiLSTM | 0.911 (154/169) | 0.716 (154/215) | 0.802 |
| Criteria2Query (CRF) | 0.902 (156/173) | 0.726 (156/215) | 0.804 |

## IV. CONCLUSIONS

In this study, we introduced a new framework CT-BERT and trained NER models to leverage BERT-based modeling for clinical trial information extraction. We studied how pre-trained BERT models may impact the NER performance. We found that BioBERT large pre-trained on PubMed/PMC works best for the clinical trial domain. We further evaluated our NER models on a benchmark data. The result showed that CT-BET outperforms baseline models including Att-BiLSTM and Criteria2Query. Collectively, CT-BERT shows significant improvement in model quality. Getting high accuracy in information extraction paves the way for automatic AI-driven clinical trial design.